%% file: main.tex
\documentclass[preprint]{vgtc}               

\graphicspath{{figures/}{pictures/}{images/}{./}} 

\usepackage{times}                     

\usepackage{tabu}                      
\usepackage{booktabs}                  
\usepackage{lipsum}                    
\usepackage{mwe}                       

\usepackage{mathptmx}                  

\usepackage{adjustbox}
\usepackage[dvipsnames,table,svgnames]{xcolor}
\usepackage{tabularray}
\newadjustboxcmd {\difftype}[1]{color = {#1} , fbox=0.5pt 1pt } 
\usepackage{multirow}
\definecolor{model}{RGB}{55, 117, 52}
\definecolor{data}{RGB}{19, 67, 112}
\definecolor{human}{RGB}{224, 142, 69}

\newcommand{\revised}[1]{\textcolor{black}{#1}}

\usepackage{tikz}

\newcommand\circleds[1]{%
  \tikz[baseline=(X.base)] 
    \node (X) [draw, shape=circle, inner sep=-0.8pt, fill=black, text=white] {\strut #1};%
}

\onlineid{1194}

\vgtccategory{Research}

\vgtcinsertpkg

\title{Towards Difficulty-Aware Analysis of Deep Neural Networks}

\author{Linhao Meng\thanks{corresponding e-mail: l.meng1@tue.nl} %
\and Stef van den Elzen %
\and Anna Vilanova}
\affiliation{\scriptsize Eindhoven University of Technology}

\teaser{
  \centering
  \includegraphics[width=\linewidth]{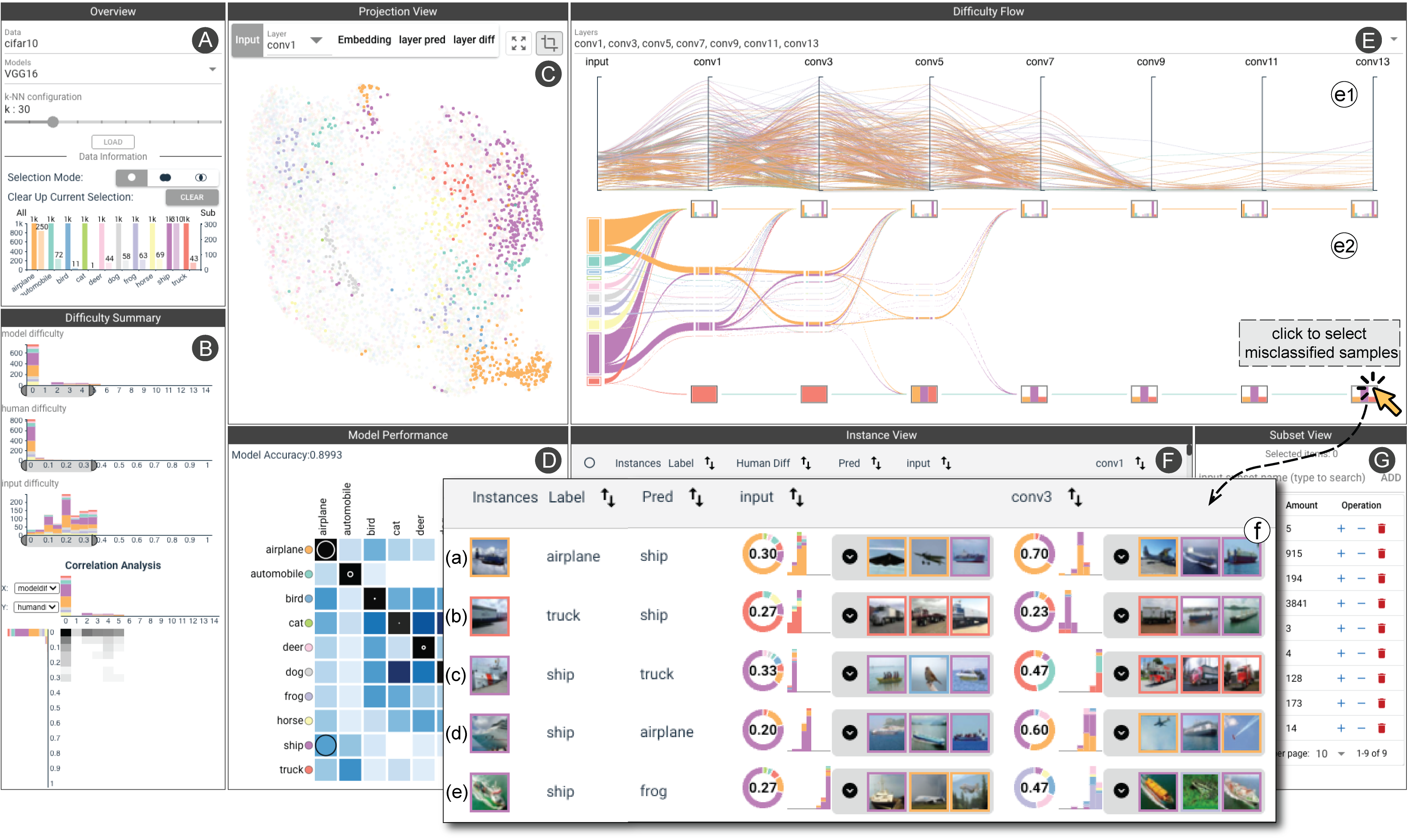}\vspace{-2mm}
  \caption{DifficultyEyes supports difficulty-aware DNN analysis from data, model, and human perspectives. Key difficulty patterns can be identified in the difficulty summary view (B) alongside the model performance view (D) showing a model prediction summary. Layer-wise difficulties are displayed in the difficulty flow view (E), while neighborhood information is available in the instance view (F). The projection view (C) enables instance similarity exploration. Selected subsets can be saved in the subset view (G).}
  \label{fig:teaser}
}

\abstract{
    Traditional instance-based model analysis focuses mainly on misclassified instances. However, this approach overlooks the varying difficulty associated with different instances. Ideally, a robust model should recognize and reflect the challenges presented by intrinsically difficult instances. It is also valuable to investigate whether the difficulty perceived by the model aligns with that perceived by humans. To address this, we propose incorporating instance difficulty into the deep neural network evaluation process, \revised{specifically for supervised classification tasks on image data.} Specifically, we consider difficulty measures from three perspectives -- \difftype{data}{data}, \difftype{model}{model}, and \difftype{human}{human} -- to facilitate comprehensive evaluation and comparison. Additionally, we develop an interactive visual tool, DifficultyEyes, to support the identification of instances of interest based on various difficulty patterns and to aid in analyzing potential data or model issues. Case studies demonstrate the effectiveness of our approach.
} 


\begin{document}



\maketitle


\input{sections/1-introduction}

\input{sections/2-relatedwork}
\input{sections/3-difficulty}
\input{sections/4-tasks}

\input{sections/5-tool}

\input{sections/6-casestudy}

\input{sections/7-conclusion}

\bibliographystyle{abbrv-doi}

\bibliography{main}
\end{document}

%% file: sections/1-introduction.tex
\section{Introduction}
Deep Neural Networks (DNNs) have demonstrated remarkable efficacy across various fields, such as image classification and natural language processing~\cite{2021deep}. Traditional evaluation metrics, such as accuracy and F1 score, provide an aggregate view of model performance, masking how models behave on individual samples. In contrast, instance-level analysis offers essential insights into model behavior, revealing systematic errors, decision boundaries, and failure modes, particularly in safety-critical or fairness-sensitive applications. Given the high-dimensional nature of data processed by DNNs, visual analysis techniques are commonly employed to support such instance-based analysis~\cite{2019deeplearningsurvey,DeepEyes,DeepVID}. Yet, most existing approaches focus solely on model outcomes, especially failure cases. This outcome-centric view is limited. For example, misclassifications can stem from fundamentally different causes -- such as ambiguous data versus overconfident errors on easy instances -- necessitating richer signals for meaningful interpretation.

\begin{table*}[tb]
\caption{Taxonomy of instance-based analysis based on instance difficulty and model correctness. }
\label{tab:diff}\vspace{-1mm}
\scriptsize%
\centering%
\SetTblrInner{
    rowsep=0pt,
    colsep=3pt,       
}
\begin{tblr}{
  width = \linewidth,
  colspec = {|c|c|c|c|c|X|},
}
\hline
\SetCell[r=2]{c} \textbf{Index} & \SetCell[c=3]{c} \textbf{Instance Difficulty}
& & &  \SetCell[r=2]{c} {\textbf{Model}\\\textbf{Correct?}}  & \SetCell[r=2]{l} \textbf{Potential Interpretation and solution} \\ \hline
& \difftype{human}{human} & \difftype{data}{data}  & \difftype{model}{model} & & \\ \hline

1a & \SetCell{brown9} low & \SetCell{brown9} low & \SetCell{brown9} low & $\surd$ & Simple, representative, and handled easily by the model — a ''clean'' case expected to be correct.\\ \hline

1b & \SetCell{brown9} low & \SetCell{brown9} low & \SetCell{brown9} low  & $\times$  & Misalignment between learned features and task semantics, or reliance on spurious patterns.\\ \hline

2a & \SetCell{brown9} low  & \SetCell{violet9} high & \SetCell{brown9} low & $\surd$ & Model generalizes well by capturing lower-level patterns that are well-aligned with the task. \\ \hline

2b & \SetCell{brown9} low  & \SetCell{violet9} high & \SetCell{brown9} low & $\times$ & Model relies on simple features but fails due to misleading similarity in raw features (e.g., background cues). \\ \hline

3a & \SetCell{brown9} low & \SetCell{brown9} low & \SetCell{violet9} high & $\surd $ & Model needs to capture subtle patterns in high-level features, possibly due to intricate internal decision boundaries or the presence of noise, outliers, or non-standard features. \\ \hline

3b & \SetCell{brown9} low & \SetCell{brown9} low & \SetCell{violet9} high & $\times$  & Model's robustness is challenged due to noise, outliers, or non-standard features, which disrupt its ability to generalize effectively. \\ \hline

4a & \SetCell{brown9} low & \SetCell{violet9} high & \SetCell{violet9} high & $\surd$  & Model generalizes well using high-level features.\\ \hline

4b & \SetCell{brown9} low & \SetCell{violet9} high & \SetCell{violet9} high & $\times$  & Model fails likely due to insufficient training on certain complex features or overfitting to less relevant patterns. Data representation may need to be enhanced or model's capacity needs to be improved.\\ \hline

5a & \SetCell{violet9} high & \SetCell{violet9} high  & low/high & $\surd$ & Likely a lucky guess or overfitting.  \\ \hline

5b & \SetCell{violet9} high & \SetCell{violet9} high  & low/high & $\times$ & Likely irreducible error.\\ \hline

6 & \SetCell{violet9} high & \SetCell{brown9} low & low/high & $\times$/$\surd$ & Ambiguous samples but representative in data (data does not reflect such ambiguity).  \\ \hline
\end{tblr}\vspace{-3mm}
\end{table*}

In this work, we enhance instance-based analysis by integrating instance difficulty into model evaluation, \revised{with a focus on supervised DNN classifiers for image data}. Specifically, we consider instance difficulty from three complementary perspectives: data, model, and human. From the data perspective, instance difficulty can be viewed as a dimension of data quality, reflecting the semantic and structural complexity of individual samples relative to the entire dataset. It is typically characterized by intrinsic properties such as similarity to other samples or the presence of ambiguous and overlapping features. From the model perspective, instance difficulty is determined by how the model processes and represents the data samples. Additionally, we incorporate human-perceived difficulty. As AI systems are increasingly deployed in critical applications, it is crucial that their decision-making processes are somewhat interpretable and aligned with human reasoning. Understanding where model and human difficulty perceptions differ or coincide helps identify potential risks and promote trustworthiness. Misalignment across these views can indicate biases in the data or limitations in the model’s learning capacity. To quantify both data and model difficulty, we adopt neighborhood-based metrics due to their interpretability. In particular, we employ Prediction Depth (PD)~\cite{2021Baldock_difficultylens}, a metric derived from neighborhood information in hidden layer embeddings, to capture model-perceived instance difficulty. \revised{This metric is applicable to DNN classifiers that produce meaningful intermediate representations, such as MLPs and CNNs.} Human-perceived difficulty is approximated using multi-annotated labels, capturing consensus and ambiguity in human judgment. Building upon selected difficulty measures, we propose a conceptual taxonomy that captures potential combinations of instance difficulties across three perspectives and then analyze the implications of these combinations with respect to model correctness, as listed in~\cref{tab:diff}. These insights highlight potential issues in data quality and model design, informing targeted interventions for data refinement or model debugging.

To operationalize this framework, we present DifficultyEyes, a visual interactive system for instance-based DNN analysis centered on multi-perspective instance difficulty. DifficultyEyes integrates coordinated visualizations to support exploration, comparison, and reasoning of difficulty patterns. To enhance understanding of model difficulty -- especially within deep architectures -- we visualize layer-wise information extracted from DNNs, illustrating how individual instances are processed across layers. We showcase the utility of this approach through two use cases. The key contributions of this work are summarized as follows:
\vspace{-2mm}
\begin{itemize}
\item A DNN analysis approach established upon the concept of instance difficulty, commencing with the presentation of instance difficulty from three perspectives -- \difftype{data}{data}, \difftype{model}{model} and \difftype{human}{human}, and extending to difficulty interpretation.
\vspace{-2mm}
\item A visual interactive tool, DifficultyEyes, designed to support the proposed workflow. This tool includes visualizations and interactions to display instance difficulties and select instances of interest for an in-depth understanding and analysis.
\end{itemize}

%% file: sections/2-relatedwork.tex
\section{Related Work}
In this section, we review previous research about instance-based model analysis and discuss various difficulty measures.
\subsection{Instance-based model analysis}
Instance-based visual analysis of models focuses on visually encoding information specific to particular instances, enabling detailed examination of instances of interest~\cite{2019deeplearningsurvey, VACCNN}. A common approach is to visualize data features alongside model results for the same instances, aiding in unraveling the connection between model input and output~\cite{DeepVID, Manifold}. Furthermore, visualizing a model's intermediate data for individual instances can unveil its inner workings and aid in diagnosing specific behaviors~\cite{RNNbow, AttentionFlows, DeepTracker, AttributionScanner}. Such information is typically high-dimensional, so dimensionality reduction is often used to project it into 2D space for analysis~\cite{2017hiddenembedding}. Given the ambiguity of overlapping points in the scatterplot, dedicated efforts have been directed towards refining visual designs that effectively present instance information~\cite{2017squares, DeepVID}. Given the substantial volume of instances, misclassified instances are often singled out and designated as instances of interest~\cite{activis,classhierarchy}. We take a distinctive approach by deriving instance difficulties from different perspectives and examining their alignment to locate instances of interest. 
\subsection{Instance difficulty measures}
We categorize instance difficulty measures in the literature into two main types. The first revolves around inherent data characteristics, encompassing factors such as similarity to other data points, noise levels, class imbalance, and the discriminative power of features~\cite{2014hardmeasures,2020Arruda_measures, 2017imbalance, 2023hardsurvey}. The computation of these measures often involves established machine learning techniques, such as k-nearest neighbors~\cite{2019KNNdifficulty,2022hardvis}. Notably, the machine learning techniques used for quantifying these measures are unrelated to the task model. This category emphasizes a focused exploration of data characteristics without being constrained by task-specific considerations. The other category concerns the behavior of the task model, thereby capturing model-perceived instance difficulty. These measures often originate from data generated during the training or prediction processes, such as parameters related to loss functions~\cite{2020superloss, 2020curriculumlearning}, backpropagated gradients~\cite{2022varianceofgradients} or ensemble behaviors~\cite{2022ILDAE}.
Existing research primarily focuses on leveraging instance difficulty during model training. For instance, it serves as metrics for data pruning~\cite{2011pruning, 2022hardvis, 2022pruning} or is included in sample weighting strategies~\cite{2023weighting, 2022weighting, 2020focalloss}, with the goal of improving model reliability and generalization. There is a relative scarcity of studies exploring instances at different difficulty levels for model evaluation~\cite{2022visualhard, 2018instancespance}.

%% file: sections/3-difficulty.tex
\section{Difficulty Quantification}\label{sec:diffquantification}
In this section, we detail our selected difficulty measures. We primarily adopt neighborhood-based metrics from the data and model perspectives, enabling interpretable explanations of difficulty through comparisons with nearby samples in the training data.
\begin{figure*}[t]
  \centering 
  \includegraphics[width=0.9 \linewidth]{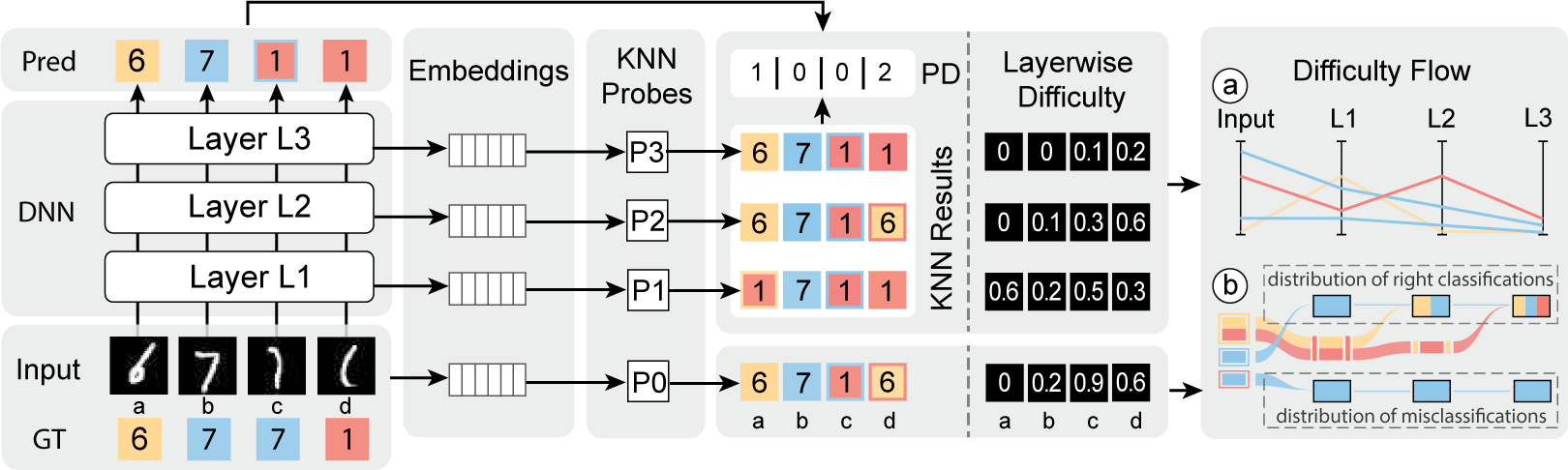}\vspace{-3mm}
  \caption{%
    Given the examined image instances (a-d), we derive layer embeddings from each layer of the DNN model. Using these embeddings, the built k-NN probes make predictions on these instances, allowing us to derive the Prediction Depth (PD). Additionally, k-Disagreeing Neighbors (kDN) scores are calculated to determine layer-wise instance difficulties. Instance difficulty from the data perspective is computed similarly. Furthermore, the k-NN results and instance difficulty values are visualized in the difficulty flow view of our tool.
  }
  \label{fig:pipeline}\vspace{-4mm}
\end{figure*}

We compute k-Disagreeing Neighbors (kDN) score~\cite{2014hardmeasures} to assess instance difficulty from the \difftype{data}{data} perspective. This score measures label inconsistency within an instance's pixel-based neighborhood, with higher kDN scores indicating greater ambiguity in the feature space. For \difftype{model}{model}-perceived instance difficulty, we use Prediction Depth (PD), a metric specifically designed for DNN classifiers. It starts with the construction of k-nearest neighbors (k-NN) classifier probes from the embeddings of the training data at specific layers of the DNN and their corresponding ground truth labels. With the constructed layer classifier probes, the computation of PD is depicted in~\cref{fig:pipeline}. Embeddings for input instances are extracted at each hidden layer, and the corresponding classifier probe is used to make predictions. PD is defined as the number of hidden layers after which the k-NN classifications consistently align with the DNN final predictions. To approximate \difftype{human}{human}-perceived difficulty, we measure the degree of disagreement among multiple human annotations relative to the given ground-truth label.

To support smooth interactive analysis, our implementation employs an approximate nearest neighbor algorithm~\cite{annoy} to accelerate nearest neighbor search, and principal component analysis~\cite{2010PCP} to reduce the dimensionality of hidden layer embeddings.

%% file: sections/4-tasks.tex
\section{Design Requirements}\label{sec:tasks}
Based on the instance difficulty measures outlined in~\cref{sec:diffquantification}, we summarize design requirements \textbf{R1-3} for our visual interactive tool. 

\textbf{R1 - Provide an overview of instance difficulty across three perspectives alongside model performance.} The tool should offer an aggregated view of instance difficulty from the \difftype{data}{data}, \difftype{model}{model}, and \difftype{human}{human} perspectives, enabling comparison to identify patterns, such as intrinsically easy instances that the model finds difficult. In line with the taxonomy defined in~\cref{tab:diff}, it should also display model correctness to support the identification of relevant patterns.

\textbf{R2 - Present detailed information to explain instance difficulty from different perspectives and interpret model reasoning with layer-wise processing.} The tool should provide visualizations to clarify instance difficulty from each perspective. For neighborhood-based difficulty measures, this might include visualizing instance similarity or neighborhood composition to aid interpretation. Additionally, layer-wise k-NN results within the DNN should be shown to help interpret the model's decisions.

\textbf{R3 - Support flexible subset selection.} Users should be able to interactively select and retain subsets of instances based on patterns identified across coordinated views, enabling focused investigation and tracing of specific data patterns.

%% file: sections/5-tool.tex
\section{DifficultyEyes}\label{sec:tool}
To support instance-based DNN analysis based on instance difficulty, we have designed and implemented a visual interactive tool, DifficultyEyes (\cref{fig:teaser}), which meets the design requirements outlined in~\cref{sec:tasks}. Once the target dataset and model are selected in the data configuration view \circleds{A}, the difficulty summary view \circleds{B} displays the distribution of instance difficulty from the three perspectives -- \difftype{data}{data}, \difftype{model}{model}, and \difftype{human}{human}, and supports a comparative analysis of two chosen perspectives. \revised{While parallel coordinate plots (PCPs) could reveal correlations among all three, we opt for heatmaps to minimize visual clutter and overlap.} A confusion matrix is shown in the model performance view \circleds{D}, fulfilling \textbf{R1}. 

To support a deeper understanding of instance difficulty from both the data and model perspectives (\textbf{R2}), we provide additional details about k-NN decisions in the difficulty flow view \circleds{E}. The computation of k-NN predictions and their associated difficulties, along with their encodings in our visualizations, is exemplified in~\cref{fig:pipeline}. Specifically, a PCP is employed to display instance difficulties from the data perspective and to connect with instance difficulties across layers, as shown in~\cref{fig:pipeline}a. To visualize the evolution of k-NN classifications, we adapt the Sankey-based design from ModelWise~\cite{modelwise}, distinguishing instances based on whether they surpass their PD, as shown in~\cref{fig:pipeline}b. Each column in the Sankey-based visualization consists of several nodes, representing k-NN predictions on the input or at a specific layer. Except for the top and bottom rows of nodes, each node represents a predicted class encoded by the color of its border or side bars. The height of each node corresponds to the number of instances predicted as the respective class. The middle bar within each node is further divided into several rectangles based on the number of instances with their actual classes. Links between columns connect rectangles corresponding to the same instances. Once instances exceed their PD on a specific layer, we know that subsequent k-NN probes will produce consistent predictions, same as final DNN predictions. Therefore, instances that exceed their PD are compressed into separate nodes above or below each column, based on whether their final predictions are correct or not. In these top and bottom nodes, bar charts are used to show their class distribution. This adaptation identifies when instances surpass their PD and conserves space to emphasize the flow of k-NN results before instances become easy to classify. 

The instance view \circleds{F} presents neighborhood information that aids interpretation of k-NN decisions in a tabular format. Given a k-NN probe and a query instance, we can query neighboring samples in the training data and retrieve their distances. In each cell of the layer columns, we present three kinds of neighborhood information: class distribution, distance distribution, and neighbor samples. The class distribution is visualized using a donut chart, with the computed difficulty score placed in the center. This score can be used for row sorting. A stacked histogram displays distance distribution between the query instance and its neighbors, which helps validate the reliability of the k-NN probe results. Images of neighboring samples are accessible through tooltips. Additionally, the projection view \circleds{C} provides an overview of the examined instances, clustering similar instances based on their feature values, embeddings of a selected layer, or overall layer-wise difficulty patterns, allowing users to select similar instances and investigate differences in their instance difficulties.

DifficultyEyes also supports flexible interactions to select instance subsets of interest across coordinated views \circleds{B} - \circleds{E} based on classes, model predictions, and difficulty information, as required in~\textbf{R3}. For example, users can brush over the difficulty distribution to filter data based on specific difficulty patterns or click on the confusion matrix to select samples based on model predictions. In addition to creating new subsets, set operations such as union and intersection are supported, allowing for flexible subset creation. Selected subsets can be saved in the subset view \circleds{G} for later reference.

%% file: sections/6-casestudy.tex
\begin{figure}[tb]
  \centering 
  \includegraphics[width=\linewidth]{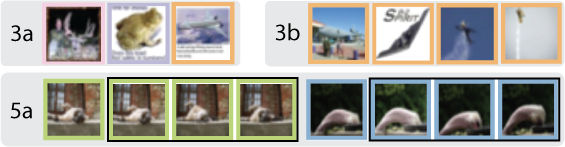}\vspace{-3mm}
  \caption{Examples following difficulty pattern 3a, 3b and 5a. The border colors of the images imply their labels as specified in~\cref{fig:teaser}A.
  }
  \label{fig:case}\vspace{-4mm}
\end{figure}

\section{Use Cases}
\textbf{Experiment Setup.} The CIFAR-10H dataset~\cite{cifar10h} extends the original CIFAR-10 dataset by adding 51 human annotations per image for the 10,000 test images. We use this dataset to calculate human-perceived instance difficulty based on annotation disagreement compared to the original labels. A VGG16 model is trained for image classification, achieving 89.93\% accuracy. Following the previously outlined methodology, we compute instance difficulties of the test images from both the model and data perspectives.\\
\textbf{Exploration of difficulty patterns from three perspectives.} Users can filter the data by brushing the plots in the difficulty summary view to focus on specific difficulty patterns. For example, as shown in~\cref{fig:teaser}B, we select instances with low difficulty values across all three perspectives -- instances that are visually simple for humans, representative in the training dataset and easily handled by the model. We observe that most of these instances belong to the airplane or ship classes, suggesting good clarity of these two classes compared to other classes. In the difficulty flow view (\cref{fig:teaser}E), although most instances are easy to classify correctly even with low-level features, we identify some instances that are consistently misclassified. By clicking on the misclassification node, we can select these samples and examine their neighborhood information in~\cref{fig:teaser}f. These instances follow pattern 1b as listed in~\cref{tab:diff}. By analyzing how the neighboring samples shift from the actual classes to the misclassified ones, we gain insights into when misalignment between learned features and task semantics occurs, as well as the potential for spurious patterns. As shown in~\cref{fig:teaser}f-(a), uncommon viewpoints (such as a direct side view) can obscure important features (like the airplane's wings), causing it to resemble a ship. Most neighboring airplane samples in the training data for input and earlier layers come from an oblique side view. Incomplete subjects, such as a truck without a cabin (b) or a ship missing part of its hull (d), also present challenges. Specifically, the very low PD of instance (b) indicates the model's high confidence in misclassifying it as a ship. Additionally, instances associated with unusual color patterns, such as a red stripe on a ship’s hull (c) or a ship with a green deck on green water (d), can lead to misclassifications since red patterns are more common in trucks, and green patterns are often associated with frogs. Our method serves for initial exploration; further evaluations are expected to be conducted to test the above assumptions about model break points. Data enhancement could be applied to improve model capability in handling these cases.

\begin{figure}[tb]
  \centering 
  \includegraphics[width=\linewidth]{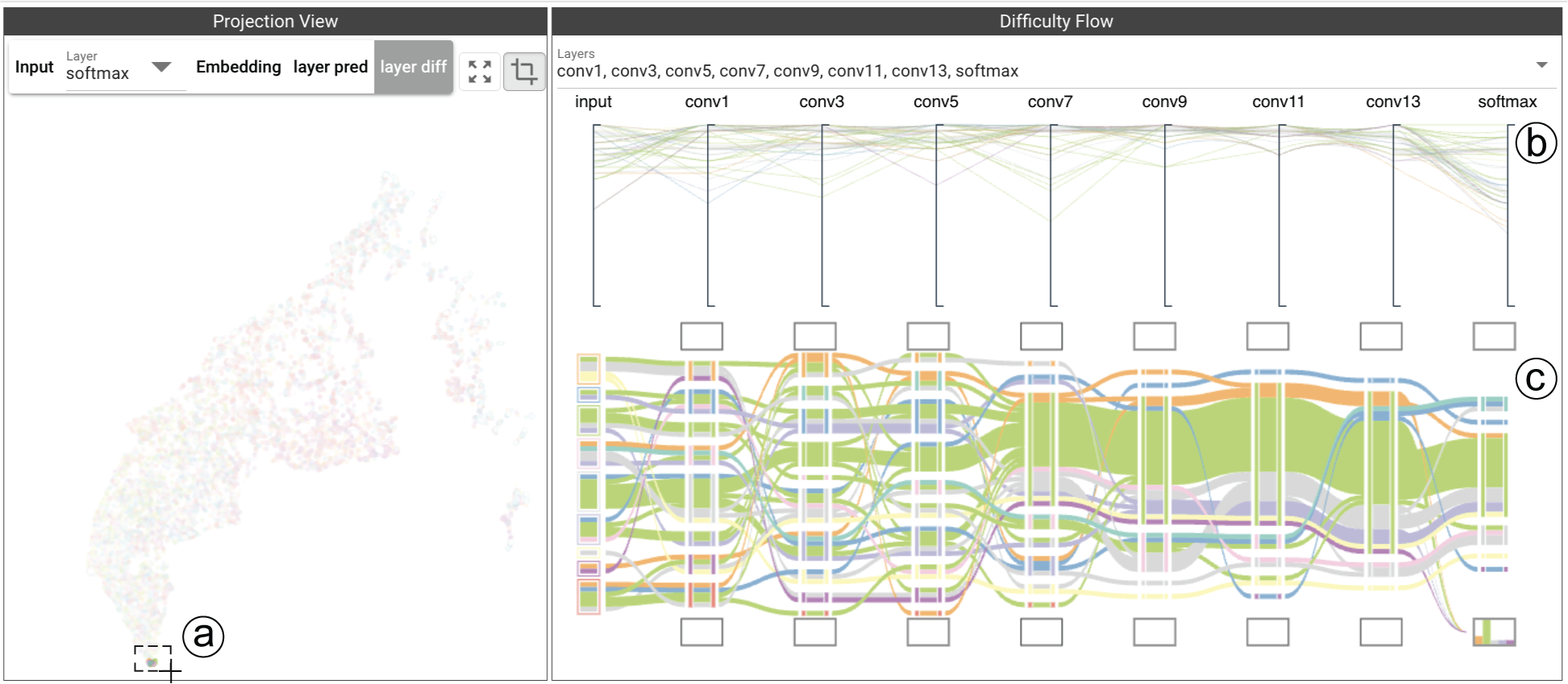}\vspace{-3mm}
  \caption{Exploration of layer-wise difficulty patterns in the model. (a) Users interactively select instances by brushing in the 2D projection view, with samples having similar layer-wise difficulty patterns are positioned closely. (b) The selected instances show high difficulty across layers. (c) However, k-NN predictions for many instances of the green class (cat) actually align with their true label.
  }
  \label{fig:case2}\vspace{-4mm}
\end{figure}

Conversely, we found instances that are visually simple for humans and well-represented in the training dataset but difficult for the model, as they only yield consistent predictions in the later layers. These instances often contain unexpected patterns that complicate the model's internal processing. For example, images with cluttered backgrounds (e.g., containing humans) or text introduce challenges. In ideal cases, the model can focus on critical features in later layers and make correct predictions (\cref{fig:case}-3a). However, non-standard features can sometimes bias the model, leading to incorrect classifications, as shown in~\cref{fig:case}-3b, where airplanes are misclassified. By examining neighboring samples, we discovered some interesting patterns. For example, in the last two images, the contrail is misinterpreted as a bird’s neck, leg, or even a branch where the bird perches. We also analyzed instances that are difficult for humans to classify. Some of these exhibit overfitting, where very similar neighboring samples in the training data (see \cref{fig:case}-5a) identified in early layers lead to correct but overly confident classifications, even when the task semantic features are unclear.\\
\textbf{Analysis of layer-wise difficulty patterns within DNNs.} By projecting instances into a 2D space based on their layer-wise difficulties, we can further analyze the patterns of difficulty progression across layers. By brushing over a small area in the projection view (\cref{fig:case2}a), we select instances that show high difficulty across all layers (\cref{fig:case2}b), indicating that the classes of their neighboring samples differ from model predictions. However, in the Sankey-based visualization view, we observe that many instances, especially those belonging to the green class (cat), show consistent k-NN predictions in the inner layers that align with their actual classes. This suggests that, while these instances have similar samples in the training data that locally yield correct results using k-NNs, the features extracted from these samples lie near the decision boundary of the DNN, or the DNN may be underfitting these cases.

%% file: sections/7-conclusion.tex
\section{Conclusion}
In this work,  we extend standard misclassification-based evaluations to focus on instance difficulty by comparing difficulty levels from three perspectives -- \difftype{data}{data}, \difftype{model}{model}, and \difftype{human}{human}. This approach allows us to examine how instances are perceived differently through various perspectives, aiding in identifying potential data or model issues. \revised{To support this analysis, we introduce an interactive visual tool designed to explore neighborhood-based instance difficulties for 2–10 image classification tasks.} Future work involves incorporating additional difficulty measures \revised{(e.g., metrics assessing other data characteristics and strategies for estimating human difficulty in datasets without multi-annotations), enhancing the scalability of our tool by extending it beyond image data and into deeper model layers, and conducting user evaluations to assess our method's effectiveness and usability.}